\documentclass[conference,compsoc]{IEEEtran}

\usepackage[english]{babel}
\usepackage{blindtext}

\usepackage{adjustbox}

\usepackage{tikz}
\usepackage{amsmath}
\usepackage{amssymb}
\usepackage{bbm}
\usepackage{enumitem}

\usepackage{url}

\usepackage{booktabs}
\renewcommand{\arraystretch}{1.2}
\usepackage{xspace}
\usepackage{cleveref}

\usepackage{listings}

\usepackage{balance}

\newcommand{\name}{Kirin\xspace}
\newcommand{\names}{Kirin's\xspace}
\newcommand{\cvr}{Cisco Virtual Router XRv9k\xspace}
\newcommand{\exabgp}{ExaBGP\xspace}

\newcommand{\eg}{e.g., \@}

\newcommand{\etal}{et al.\xspace}
\newcommand{\one}{(1)~}
\newcommand{\two}{(2)~}
\newcommand{\three}{(3)~}

\newcommand{\dbsub}[1]{#1}

\newcounter{takeaway_cnt}
\newcommand{\takeaway}[1]{\stepcounter{takeaway_cnt}\textbf{Takeaway \arabic{takeaway_cnt}:} \emph{#1}}

\usepackage{pifont}%

\begin{document}
\date{}
\title{\name: Hitting the Internet with Millions of Distributed IPv6 Announcements}

\author{\IEEEauthorblockN{Lars Prehn}
\IEEEauthorblockA{Max Planck Institute for Informatics\\
lprehn@mpi-inf.mpg.de}
\and
\IEEEauthorblockN{Pawel Foremski}
\IEEEauthorblockA{IITiS PAN\\
pjf@iitis.pl}
\and
\IEEEauthorblockN{Oliver Gasser}
\IEEEauthorblockA{Max Planck Institute for Informatics\\
oliver.gasser@mpi-inf.mpg.de}}

\maketitle

\begin{abstract}
The Internet is a critical resource in the day-to-day life of billions of users. To support the growing number of users and their increasing demands, operators have to continuously scale their network footprint---\eg by joining Internet Exchange Points---and adopt relevant technologies---such as IPv6.
IPv6, however, has a vastly larger address space compared to its predecessor, which allows for new kinds of attacks on the Internet routing infrastructure.

In this paper, we revisit prefix de-aggregation attacks in the light of these two changes and introduce \name---an advanced BGP prefix de-aggregation attack that sources millions of IPv6 routes and distributes them via thousands of sessions across various IXPs to overflow the memory of border routers within thousands of remote ASes. Kirin's highly distributed nature allows it to bypass traditional route-flooding defense mechanisms, such as per-session prefix limits or route flap damping.

We analyze the theoretical feasibility of the attack by formulating it as a Integer Linear Programming problem, test for practical hurdles by deploying the infrastructure required to perform a small-scale Kirin attack using 4 IXPs, and validate our assumptions via BGP data analysis, real-world measurements, and router testbed experiments. Despite its low deployment cost, we find Kirin capable of injecting lethal amounts of IPv6 routes in the routers of thousands of ASes. 
\end{abstract}

\section{Introduction}

The Internet is an indispensable resource for communication, trade, commerce, education, and entertainment in today's world.
Over the past years, the Internet has become more and more important in people's everyday life.
Moreover, the reliance of many societies on the Internet has only increased with the COVID-19 pandemic \cite{feldmann2021year,lutu2020characterization,bottger2020internet,bronzino2021mapping,bouwman2022helping}.

To counter IP address exhaustion among other things, the IPv6 protocol was designed more than 20 years ago \cite{rfc2460}.
Although IPv6 usage was low initially, more and more websites, services, and networks are now using IPv6.
Around 20\% of all websites are IPv6-ready~\cite{W3T2021IPv6}, a third of all Autonomous Systems (ASes) announce IPv6 routes~\cite{Geoff2021IPv6}, and around 40\% of Google users globally access the website via IPv6, with some countries reaching a deployment of more than 60\% \cite{google-ipv6}.

However, the additional capabilities provided by IPv6 come with new threats: \eg
targeted probes can find home routers in the vast IPv6 address space \cite{rye2020discovering,gasser2018clusters};
privacy mechanisms can be defeated and devices can be tracked over time \cite{rye2021follow};
even a single device using legacy IPv6 addressing can foil all privacy extension and prefix rotation efforts \cite{saidi2022one}.
In addition to these attacks on the data plane, IPv6 also introduces new challenges for the control plane.
Its vast address space raises questions about the scalability of the Internet's standard interdomain routing protocol: the Border Gateway Protocol (BGP).
Some large networks own /19 IPv6 prefixes, each of which contain \emph{half a billion} possible /48 subprefixes that reliably propagate over BGP.
As routers have a limited amount of memory available, such a large number of IPv6 prefixes would exhaust the memory of many routers deployed today on the Internet.

In this paper, we introduce and analyze a BGP flooding attack named \name---standing for \textbf{K}illing \textbf{I}nternet \textbf{R}outers in \textbf{I}Pv6 \textbf{N}etworks---that overcomes traditional protection mechanisms (\eg per-session prefix limits and route flap damping) by originating millions of unique IPv6 routes distributed across many BGP sessions. More specifically, we make the following key contributions:

\begin{itemize}[leftmargin=*]

\item \textbf{Kirin:} We describe a BGP flooding attack, \name, its threat model and recent technology trends that enable it (cf. \Cref{sec:overview}).

\item \textbf{Theoretical Feasibility Analysis:} We combine real-world data with an Integer Linear Programming definition of our attack to theoretically analyze its feasibility wrt. \one required IXP presence, \two required sessions at each IXP, and \three the resulting ASes that are affected. We show that \name is not only theoretically feasible, but can already affect ASes globally when connecting to just 20 transit providers at 25 IXP peering LANs (cf. \Cref{sec:theory}).

\item \textbf{Router Testbed Evaluation:} We test the effects of \name on one virtual and one physical router from different vendors. We find that we can exceed router memory already with 109k specially crafted IPv6 prefix announcements (cf. \Cref{sec:hwtest}).
    
\item \textbf{Real-world Experiment:} To demonstrate how \names requirements can be met in the real world, we deploy the infrastructure needed to perform a small-scale \name attack. Our testbed was built from scratch and fully functional in a few weeks and cost only 500 EUR (cf. \Cref{sec:exp:resources}).

\item \textbf{BGP Testbed Validation:} We validate our assumptions on how routes propagate using BGP data analysis and real-world experiments from our own and the PEERING testbed (cf. \Cref{sec:exp:redistribution}).

\item \textbf{Defense Mechanisms \& Notification:} We extensively discuss possible defense mechanisms (cf. \Cref{sec:discussion}) and lay-out our plan for a vulnerability notification campaign (cf. \Cref{sec:ethics}).

\end{itemize}

\section{Background}\label{sec:bg}

BGP is the standard interdomain routing protocol, where Autonomous Systems (ASes; groupings of routers) announce and redistribute reachability information between each other according to certain routing policies~\cite{rfc7454}. When an AS receives an announcement, it usually consists of an IP prefix and a path of ASes to traverse; the term \emph{route} thus is used to refer to a prefix-path pair.

\noindent\textbf{Routers.} Routers within ASes establish dedicated TCP sessions over which BGP is run between them. For the IPv4 and IPv6 protocols separately, each router holds a routing information base (RIB) that contains all currently active routes. For each prefix, a router determines its current best-path from all alternatives, and then installs the best-path's next-hop in its forwarding information base (FIB). The FIB is then used to quickly retrieve the next-hop to which the router forwards a packet. To allow a router to achieve high throughput, the FIB is often stored in expensive, specialized memory formats such as TCAM or DRAM, which are optimized to quickly perform longest-prefix-match operations. This specialized memory is often a scarce resource due to its high cost, a fact that has previously been exploited for theoretical stress attacks~\cite{deng2010evaluating}.

\noindent\textbf{Route Propagation in Theory.}
Once a router determines the best-path for a given prefix, it may redistribute the new route to its BGP neighbors. 
Whether a route is redistributed to a certain neighbor is determined by applying a chain of egress filter rules; the sum of these rules commonly expresses more abstract policies that represent a network's business incentives. 
In 2001, Gao and Rexford first categorized the business relationships between ASes and identified the consequential redistribution patterns~\cite{gao2001stable}. The Gao-Rexford model describes three types of relationships: \one transit relationships in which a customer pays a transit provider to forward traffic, \two peering relationships in which two ASes achieve mutual benefits by forwarding traffic for one another at no cost, and \three sibling relationships in which two ASes appear as two logically separate AS numbers but are operated by the same organization and hence produce ``arbitrary'' redistribution patterns.
Based on these relationship categories, ASes only redistribute routes that provide monetary benefit. While ASes would redistribute routes they received from their customers to all other neighbors (as the customer ultimately pays for the delivered traffic), they would not forward routes that they received from peers to other peers or transit providers (as the peer would not pay for resulting maintenance or transit costs). 

\noindent\textbf{Route Propagation in Practice.}
While these abstract relationship models still hold today~\cite{gill2013survey,luckie2013relationships,jin2020toposcope}, they are partially superseded by more nuanced relationships~\cite{giotsas2014inferring,drPeering2022Playbook}, e.g., partial transit (i.e., for a limited set of routes), paid peering (i.e., one AS pays the other a small fee to access routes towards its customer cone---the set of all direct and indirect customers), or hybrid relationships, where the actual relationship between two ASes depends on the physical location. Besides business relationships, the propagation behavior of an AS can be influenced by various factors including \one route reputation---some ASes may filter and ignore routes if they or their originating AS appear in block lists \cite{SPAMHAUS2022DROP,Cymru2022Bogon,AbuseIPDB2022List}---\two aggregation strategy---ASes may aggregate routes before redistribution to limit routing table growth~\cite{kalogiros2009understanding, le2011route} or provide customers with (partial) default routes~\cite{rodday2021deployment}---or \three remote signaling where, e.g., customers instruct their providers to redistribute a route in a certain way using BGP Communities~\cite{streibelt2018bgp,birge2019sico}.

\noindent\textbf{Propagation Timing.} There are also factors that determine \emph{when} a router propagates a route. Many ASes configure a Minimal Route Advertisement Interval (MRAI) during which announcements are aggregated; after the MRAI timer expires, only the currently active best-path is propagated, which substantially reduces the number of propagated updates due to route flapping\footnote{i.e., routes that generate many update messages as they rapidly shift between two or more configurations.} ~\cite{gill2012effect,fabrikant2011there}. Another widely deployed approach which influences the propagation time is Route Flap Damping (RFD)~\cite{gray2020bgp,pelsser2011route}. An RFD-enabled BGP session keeps a penalty counter for each prefix. The counter is incremented for each received update and decremented at fixed time intervals. If the counter exceeds a ``suppress'' threshold, the router starts to dampen the prefix, i.e., it withdraws it from all its peers and no longer redistributes updates for it. It remains in this state until the counter has decreased to some ``reuse'' threshold, after which it starts to redistribute the prefix again.

\noindent\textbf{Path Exploration.} A router may enter a ``path exploration'' period once it receives a withdraw. When the origin AS entirely withdraws a prefix, a remote AS receives the withdraw messages from different paths spread across a certain time window---a result of the propagation timings of the routers along a path. If a router knows multiple paths for a prefix and it receives a withdraw for its current best-path first, then it chooses some other path as its new best-path and generates an update that reflects the change. If a router knows $N$ paths for a prefix, it may repeat this cycle up to $N-1$ times (in the worst case) before it finally redistributes the withdraw message: i.e., it ``explores'' potentially all of the other available paths before it fully withdraws the prefix. Path exploration is present in most (if not all) active route propagation experiments and has been studied extensively~\cite{mao2003bgp,alabdulkreem2014using}.

\noindent\textbf{Internet eXchange Points (IXPs).} Over the last decade, peering has increasingly gained importance~\cite{bottger2018elusive}. IXPs allow their members to cost-efficiently establish peering sessions with other members on top of their existing peering LANs, i.e., layer 2 switching infrastructures that are bound to specific geographic locations~\cite{ager2012anatomy}. Many IXPs also provide route servers to further facilitate peering: using a single BGP session, an IXP member can exchange routes with all other (often 500 or more) ASes that are connected to the route server~\cite{richter2014peering}. As of 2022, there are more than 650 active\footnote{We consider only IXPs with operational status ``active'' and at least two participants.} IXPs worldwide~\cite{PCH2022ixpdir}. Some of these IXPs provide access to more than 1000 potential peering partners and routes for more than half of the Internet~\cite{bottger2018elusive,prehn2022peering}.
These reachability benefits provided by IXPs also make remote participation attractive. Nowadays, ``remote peering,'' i.e., connecting to a peering LAN via some layer 2 connectivity provider, has become the norm rather than an exception ~\cite{castro2014remote,nomikos2018peer,mazzola2022latency}.

\noindent\textbf{Topology Blindness.} While IXPs are highly popular and have been shown to enable hundreds of thousand of interconnections, most of these interconnections are invisible to the existing BGP monitoring platforms~\cite{ager2012anatomy,prehn2022peering}. While these platforms, in total, operate 50+ route collectors that receive and dump routing updates from 600+ feeding ASes, they, in general, miss many peering links as those often do not propagate to any feeding AS~\cite{ager2012anatomy,arnold2020cloud,giotsas2013inferring,oliveira2009completeness}.

\noindent\textbf{Route Aggregation \& Filtering.} To reduce their routing table size, some ASes perform route aggregation, i.e., they summarize multiple more-specific routes into a single less-specific route and only propagate this summary route~\cite{kalogiros2009understanding,le2011route,gamba2017bgp}. Besides aggregating routes, operators often configure their routers to ignore specific types of routes. Especially routes towards small amounts of address space (i.e., those with CIDR sizes more specific than /24 and /48 for IPv4 and IPv6, respectively) are very commonly filtered~\cite{sediqi2022hyper,ripe2019hyperspecifics}.

\subsection{Related work}

While the option to de-aggregate a prefix has been well-known in the operator community for multiple decades, academic focus on the issue is limited.

Chang \etal experimentally investigated the response of 3 commercial grade routers to large BGP routing tables in 2002~\cite{chang2002empirical}. The authors found significant differences in how routers respond and highlighted that the BGP graceful restart capability could alleviate the effects of BGP malfunctions on IP routing. A deliberate attack and its impact on the Internet are outside the scope of that paper. Yet, similar to Ceasar \etal \cite{caesar2005bgp}, the authors advocate for the use of prefix limits on BGP sessions. The operator community largely shares this sentiment as prefix de-aggregation often exacerbates the impact of route leaks~\cite{leaks2004anatomy,leaks2022cloudflare}.

In 2013, Schuchard \etal first described the concept of prefix de-aggregation attack for IPv4~\cite{schuchard2013peer}. While they describe the same underlying idea, in comparison with \name, the paper does not consider various practical details: \one they assume that the attack is executed by major transit networks with rich peering fabrics; \two they assume that an AS can obtain enough address space via squatting (illegitimately announcing unused address space) and that filters against squatting are negligibly deployed; and \three they assume that typical max prefix limits range between tens of thousands of prefixes and the full routing table size. While our work builds upon the same simple idea, it actively addresses these real-world issues, ultimately rendering the attack practically feasible: \one based on discussions with network operators, we assume that prefix limits are widely deployed and usually range between hundreds to a few thousand prefixes on peering sessions; \two we leverage IPv6 as an enabler to source millions of legitimately allocated---and hence unfiltered and even RPKI-valid---prefixes; and \three we leverage remote peering providers, VPS providers, and IXPs to assemble thousands of sessions allowing arbitrary actors to execute \name at a minimal cost. Thus, besides a theoretical feasibility analysis, we evaluate the interlinking parts of our improved attack model in practice and on real-world Internet data.
\section{\name: Overview}\label{sec:overview}

In essence, the \name attack is simple and ostensibly obvious: the attacker introduces enough new IP routes to overflow the FIB and/or RIB of the BGP routers within victim ASes.
After that, the attacker simultaneously withdraws all previously established routes, which triggers the path-hunting phenomenon that leads to a flood of update messages that impact the performance of routers.

The idea that routers may crash due to memory constraints is not new: many operators already reported crashed routers when the IPv4 routing table reached 512K and 768K routes~\cite{Emile2021512K,Emile2021768K}.
Nowadays, high-end devices from major router vendors support \textasciitilde2--4M routes in total in their FIB: Cisco's Catalyst 8200 and 8500 platforms can store between 800k and 4M routes (depending on the exact model and its respective DRAM storage~\cite{Cisco2022FIB200,Cisco2022FIB500}), Arista's FlexRoute Engine can store up to 2.5M total routes~\cite{Arista2022FIB}, and Juniper's PTX10001 platform can handle 2M total routes~\cite{Juniper2022FIB}. 

However, it is the new context and availability of new methods that we believe re-enable a well-known attack to be successfully executed on the Internet today, by anyone, and with a limited budget.
Although there are various roadblocks built into the routing ecosystem to prevent the exploitation of the FIB/RIB overflow issue,
\name uses a set of observations and tricks to maneuver the existing roadblocks.

\subsection{Threat Model}\label{ssec:threat}
Our threat model, which was already introduced in a similar form by Schuchard et al. \cite{schuchard2013peer}, focuses on highly connected ASes with legitimate BGP speakers that act maliciously. The goal of our adversary is to fill the FIB or RIB within a remote router to the point where it fully exhausts the available memory using millions of prefix announcements. Hereby, the adversarial AS is not limited to transit ASes; as we demonstrate in §\ref{sec:theory} that even stub ASes are capable of reaching this goal. In fact, we show in §\ref{sec:exp:resources} that an adversary can start without any resources or infrastructure and yet is able to perform prefix de-aggregation attacks within less than a month and at a cost bearable for individuals. Notably, an AS may either intentionally decide to become an adversary (and explicitly assemble the required infrastructure) or may be forced in this role by an outside entity that compromised various BGP routers or a global route controller. 

While an adversary's router can only send BGP messages to the direct neighbors it established sessions with, it relies on those genuine peers to redistribute these messages according to common BGP policies. Further, our adversary may potentially ignore best common routing practices, yet must assume that all other ASes may implement them. 

\subsection{Enablers}\label{ssec:enablers}

\textbf{IPv6.} IPv6 addressing space is so much bigger than IPv4, that instead of assigning 1 IP address to 1 end-user---or even many more end-users through Network Address Translation (NAT)---in IPv6 end-users are typically assigned /64 prefixes each.
As a consequence, Internet operators also handle much bigger IP prefixes, e.g., ARIN's allocation policy states that an ISP should never receive less than a /32 prefix allocation~\cite{arinSize}.
Given that the smallest IPv6 prefix that reliably propagates over BGP is /48 \cite{ripe2019hyperspecifics,sediqi2022hyper,prehn2020wells}, potential bad actors could split their typical IPv6 prefix into \emph{much} more subnets compared with their typical IPv4 prefix. 
Splitting a /29 IPv6 prefix is enough to inject 1M unique and valid routes into the global routing table. Note that these sub-prefixes can overlap: \eg a /46 prefix can source 7 routes in total (1x /46, 2x /47, and 4x /48). In general, if $C$ is the difference between the smallest propagating CIDR size (typically 48) and the parent prefix length, an attacker can source up to $2^{C+1}-1$ unique routes. 

\noindent\textbf{Ineffective Route Aggregation.} Given that we source all prefixes from the same continuous address space, a wide deployment of aggregation would nullify \names attack potential. To overcome this challenge, \name only announces non-aggregatable prefix combinations to each neighbor and may also alternate its origin AS. Please note that the use of small, non-aggregated IPv6 prefixes is already common, and that the average prefix length is increasing over time~\cite{Geoff2022aggr, Geoff2022plen}.

\noindent\textbf{Per-Session Max-Prefix Limits.} The most commonly recommended approach to prevent the announcements of many routes is to set a maximum number of accepted prefixes for each BGP session. 
Upon hitting this limit, the session may produce a warning, might be capped---i.e., stop accepting updates for new prefixes yet keep updating existing ones---or can be dropped entirely~\cite{CISCO2022Maxpreflim}.
Because this approach requires only per-session state, it is simple to implement and requires no cooperation---two key factors that pushed today's wide deployment. 
\name attempts to respect per-session limits by distributing a dedicated set of prefixes to each of \emph{many} BGP sessions: no single prefix is shared between any two sessions.
Using this strategy transforms the goal of announcing millions of routes into a session-hunting challenge. 
We further explore this relation theoretically and experimentally in Sections \ref{sec:theory} and \ref{sec:experiments}, respectively.
Moreover, during our experiments we find IP transit and IXP operators to be permissive about increasing the prefix limits when inquired. One major transit provider stated they do not impose prefix limits on IP transit links; another stated they allow the limit that we set ourselves in the Internet Routing Registry (IRR).

\noindent\textbf{Accessible Internet resources.} It is relatively easy to obtain an AS number and a large IPv6 prefix valid for use in the global routing system.
A quick and relatively cheap way is to use services of a \emph{sponsoring} LIR, who proxies a request for resources to one of the 5 RIRs (\eg \cite{securebit-offer}).
LIR operators can \emph{lease} their allocated IP space, e.g., some offer /29 prefixes with a 48h free trial \cite{rapidseedbox}, which is enough to launch \name.
Another essentially free (yet illegal) method for malicious attackers could be \emph{squatting}, a method in which non-announced Internet resources allocated to an unrelated organization are used~\cite{nemmi2021parallel}.
Finally, it is also possible to become a regular LIR and gain direct access to legit and large IPv6 allocations. 
For example, as of 2022, becoming a RIPE member costs under 2500 EUR and allows for /29 IPv6 allocations without providing any justification \cite{ripe-membership,ripe-request-ipv6}.

\noindent\textbf{Instant and cheap BGP peering.} It is no longer true that in order to establish a BGP session neighboring networks must be physically connected \cite{nomikos2018peer}.
Remote peering at IXPs is an established reality, and a recent study found that already over 10\% of members of major IXPs are remote \cite{mazzola2022latency}.
Commercial services allow for instantly establishing peering links with dozens of significant IXPs, cloud operators, and data centers \cite{ixreach,retn,megaport}.
Furthermore, prompt provision of VMs with IXP peering sessions has never been easier: \eg a VM with NL-IX peering could cost under 30 EUR per month \cite{ifog}, and a VM with BGP IP transit could cost just \emph{a few} USD per month~\cite{vultr}.
Moreover, while carrying our experiments for this paper, we found it is easy to obtain \emph{free} IPv6 transit---foremost from Hurricane Electric (HE), a major Internet operator, who actively seeks to establish bi-lateral peering sessions with new IXP members.
We also inquired a few other major operators and found the cost of a BGP peering port with IP transit would cost around 100--300 USD per month, depending on location and bandwidth.

\noindent\textbf{Circumventable Filtering.} While it is hard to enter millions of route-objects into IRR databases, many providers nowadays also accept routes with valid ROAs. As ROA entries allow for CIDR ranges, an adversary may enter a single ROA with CIDR sizes /29--/48, wait for it to propagate, and then would pass, e.g., the route filtering checks of HE.~\cite{HE2022Checks}.

\subsection{Collateral Damage via Path Hunting}

While \name itself mainly fills the FIB/RIB of victim ASes, it does so by announcing millions of routes globally that, at some point, need to be withdrawn from the Internet again. If a global route gets fully withdrawn, the path-hunting phenomenon may produces a burst of updates (see §\ref{sec:bg} for details). 

Given that \name triggers this phenomenon simultaneously for millions of prefixes, it ``accidentally'' generates a distributed update flooding attack. Given that some ASes use route flap damping to ignore these announcements and stop the redistribution, it is hard to provide realistic estimates on the number of produced updates at each AS. Hence, we leave the analysis of collateral damage as future work and focus on \names main idea: propagation of millions of prefixes via thousands of distributed sessions.  %
\section{Theoretical Feasibility Analysis}~\label{sec:theory}

In this section, we theoretically analyze \names feasibility. We consider two different scenarios: \one the adversary obtains (potentially costly) transit from a few providers and \two the adversary obtains as many (virtually cost-free) bi-lateral and multi-lateral peers as possible. While, in reality, an adversary may use both of these scenarios simultaneously, examining them independently allows us to keep our analysis reasonably simple while still obtaining deep insights into \names cost-benefit trade-off. Further, we assume that an adversary only establishes a single (virtual) port via a single method and service provider at each peering LAN. 

We start this section by clearly stating the assumptions we make about route redistribution (§~\ref{sec:theory:assump}) and the data sources that we build our analysis upon (§~\ref{sec:theory:data}). We then define the cost-benefit trade-offs for the first and second scenario as ILP problems (§~\ref{sec:theory:form:transit} and §~\ref{sec:theory:form:peering}) and finally discuss our analysis results (§~\ref{sec:theory:analysis}).

\subsection{Assumptions \& Definitions}\label{sec:theory:assump}

\textbf{Routing Policies and Assumptions.} The policies that underpin today's inter-domain routing mostly follow economical incentives~\cite{anwar2015investigating}. In particular, we assume that:

\textbf{1)} If an AS receives a route from a customer, it forwards the route to all neighbors.

\textbf{2)} If an AS receives a route from a settlement-free peer or a provider, it forwards the route to customers only.

\textbf{3)} An AS will always forward a route by the above rules to maximize its economical gain.

The first and second assumptions are known as the Gao-Rexford redistribution model~\cite{gao2001stable} and have been standard assumptions for more than two decades in the field of AS relationship inference~\cite{luckie2013relationships,jin2019stable,jin2020toposcope,feng2019unari,giotsas2014inferring}; the third assumption has frequently (yet usually implicitly) been used for simulating route propagation, e.g., ~\cite{karlin2006pretty,zhang2007practical,morillo2021rov++}. Notably, these assumptions do not always capture the real-world behavior of all ASes perfectly (see, e.g., complex relationships~\cite{giotsas2014inferring} or non-economic incentives~\cite{gill2013survey}), yet their frequent appearance in the related literature renders them as reasonable abstractions. Based on these assumptions, Luckie et al. introduced the notion of the customer cone, i.e., the set of all direct and indirect customers of an AS~\cite{luckie2013relationships}. While they introduced multiple methods to calculate this set, we choose the one that only uses routes the AS forwarded to its peers and providers, as it yields more stable and realistic results. By recursively applying our three assumptions, one arrives at the high-level statements: \one routes sent to a peer will eventually reach all ASes in the peer's customer cone and \two routes sent to a transit provider will eventually reach all\footnote{Notably, there are certain situations in which a route does not propagate, e.g., because it is filtered or because certain ASes only want to have a default route from their providers.} ASes globally.

\begin{figure}[h!tb]
    \centering
    \includegraphics[width=.66\linewidth]{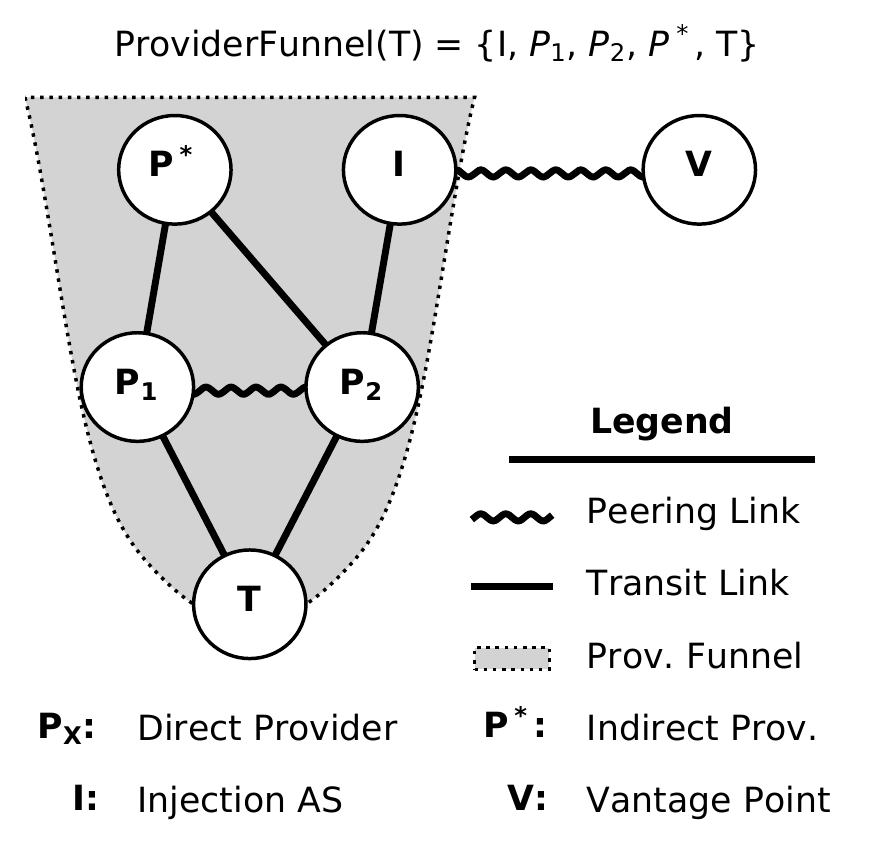}
    \caption{Provider funnel example.}
    \label{fig:fd_example}
\end{figure}
\vspace{-0.5em}

\noindent\noindent\textbf{Provider Funnel \& Funneling Degree.} In this paper, we introduce the concept of \emph{provider funnel} $PF_T$ as the set of all recursively added providers for a given target AS $T$. We use the example in Figure \ref{fig:fd_example} to further illustrate this concept. In our example, $T$ is multi-homed to two direct providers---$P_1$ and $P_2$. Neither $P_1$ nor $P_2$ are Tier1 ASes, so they also rely on different transit providers $P^*$ and $I$ to reach certain parts of the Internet. When $P^*$ announces a route to $P_1$, $P_1$ likely forwards this route to $T$. Even though $P^*$ and $T$ share no direct connection, $P^*$ is an indirect provider of $T$.

When executing \name, our vantage point $V$ has connections to ASes within $T$'s provider funnel. As these ASes redistribute our routes so they ultimately reach $T$, we call them \emph{injection ASes}. Moreover, as $V$ might maintain multiple BGP sessions to $I$ (e.g., at different IXPs), we further define an \emph{injection session} as a unique BGP session to an injection AS.

Finally, we call the number of ASes in $PF_T$ as the \emph{funneling degree} of T and denote it as $FD_T$. Note that we include $T$ in its provider funnel, i.e., $PF_T = \{P_1, P_2, P^*, I, T\}$. We use the term \emph{restricted funneling degree} $FD_{T}^{S}$ to refer to the size of the provider funnel when only considering ASes in $S$, i.e., $FD_{T}^{S} = |PF_T \cap S|$.

\subsection{Data Sources \& Processing}\label{sec:theory:data}
We estimate funneling degrees using two inputs: \one the number of sessions that each AS has with each peering LAN and \two the provider funnel for each AS.

\noindent\textbf{Estimating Peering LAN Sessions.} On 2022-09-09, we generated a snapshot of EURO-IX's IXP database~\cite{EUROIX2022IXPDB}. We further obtained a PeeringDB snapshot for that day from CAIDA's daily archive~\cite{CAIDA2022PDB}. While the EURO-IX data set does not contain a direct reference to the IXP, it contains the PeeringDB identifier for each co-location facility, which allowed us to merge the \texttt{(peering LAN, ASN, IPv6 address)} triplets we extracted from both data sources. The obtained data describes 24k sessions via 725 peering LANs. 

\noindent\textbf{Estimating IPv6 Provider Funnels.} While CAIDA publishes provider-peer-determined customer cone files on a monthly basis (available at~\cite{CAIDA2022ASREL}), this data set comes with two problems: \one it it not available for the IPv6 routing ecosystem and \two it only uses data from public route collectors which miss significant portions of the AS topology. Hence, we generate this data set (and most of the required tooling) from scratch. 

We first extract all IPv6 routes from public route collector data via BGPStream on 2022-09-09 (including routes from all RIB snapshots and update messages). Next, we add routes from 130 IPv6 route servers of 11 IXPs---e.g., DE-CIX, LINX, and IX.br---including both primary and (potentially multiple) secondary servers. All of these route servers provide a public Alice-lg looking glass utility~\cite{DECIX2022ALICELG} that has a back-end API allowing for obtaining all IPv6 routes received from their peers. We automated the querying process and obtained the IPv6 routes of all route servers throughout 2022-09-09.  

To estimate AS relationships, we utilize the publicly available ASRank script~\cite{CAIDA2022ASREL}. We modify the script to tailor it towards the IPv6 ecosystem~\cite{giotsas2015ipv6}. We use the previously collected IPv6 routes and a list of route server ASNs---that we obtained by selecting ASNs with the ``Route Server'' network type within our PeeringDB snapshot---as input to the modified ASRank script, which leads to the inference of 247K peering links and 32K transit links. Finally, we  convert the IPv6 paths and business relationships into peer-provider-determined customer cones \cite{muller2019challenges}. To calculate provider funnels, we inverted these customer cones, i.e., we checked for each AS in which other AS' customer cone it appears.

\subsection{ILP Formulation: Transit Scenario}\label{sec:theory:form:transit}
Now that we obtained the required data sets, we can formalize \names resource needs and attack potential. In our first scenario, we assume that the adversary chooses multiple transit providers and then joins peering LANs to establish additional sessions with the chosen providers. As discussed in §~\ref{sec:theory:assump}, we assume that routes announced to a transit provider propagate globally. As every prefix reaches each AS globally, we can focus on the number of sessions that can be obtained by using $P_{max}$ providers and connecting to $L_{max}$ peering LANs. 

\noindent\textbf{Sets.} Let $A$ be the set of all IPv6-enabled ASes and $L$ be the set of all peering LANs.

\noindent\textbf{Parameters.} Let $\omega_{a, l}$ denote the number of unique sessions that can be established with AS $a \in A$ at peering LAN $l \in L$. We can then build the following session matrix:

$$ S = 
 \begin{pmatrix}
  \omega_{a_1, l_1}  & \omega_{a_2, l_1}  & \cdots & \omega_{a_{|A|}, l_1}\\
  \omega_{a_1, l_2}  & \omega_{a_2, l_2}  & \cdots & \omega_{a_{|A|}, l_2}\\
  \vdots            & \vdots            & \ddots & \vdots  \\
  \omega_{a_1, l_{|L|}}  & \omega_{a_2, l_{|L|}}  & \cdots & \omega_{a_{|A|}, l_{|L|}}\\
 \end{pmatrix}
$$

We further provide the parameters $L_{max} \in \mathbb{N}$ and $P_{max} \in \mathbb{N}$ that reflect the maximum number of peering LANs and providers that can be chosen. 

\noindent\textbf{Variables.} We first introduce a binary decision matrix $D$ that contains a binary decision variable $d_{a, l}$ for each $\omega_{a, l}$ that denotes whether provider $a \in A$ at peering LAN $l \in L$ is part of the solution. Further we introduce two sets of binary decision variables that help us to realize our constraints: $CL$ contains a variables $cl_l$ for each $l \in L$ that determines whether the adversary has to connect to peering LAN $l$ while $CP$ contains a variable $cp_a$ for each $a \in A$ that determines whether $a$ is chosen as a transit provider 

\noindent\textbf{ILP Problem Formulation.} Given $S$, $L_{max}$, and $P_{max}$, our goal is to chose a set of providers and a set of LANs such that we can obtain the maximum number of sessions, i.e., 

\begin{equation*}
\begin{array}{lll}
\text{maximize} & \displaystyle\sum\limits_{l \in L}\displaystyle\sum\limits_{a \in A} \omega_{a, l} * d_{a, l} &
\end{array}
\end{equation*}

To ensure that only $L_{max}$ LANs and $P_{max}$ ASes are chosen, we add the following two constraints:

\begin{equation*}
\begin{array}{lll}
\text{wrt.} & \displaystyle\sum\limits_{l \in L} cl_l \leq L_{max} &\\
            & \displaystyle\sum\limits_{a \in A} cp_a \leq P_{max} &
\end{array}
\end{equation*}

Next, we need to make sure that $d_{a, l}$ is always 0 whenever either $cl_l$ or $cp_a$ are 0---if a LAN/AS is not chosen, its entire line/row should only contain zeros. If both, $cl_l$ and $cp_a$, are set to 1, we want $\omega_{a, l}$ to be arbitrarily large (the more sessions can be obtained, the better). To represent this circumstance we introduce a ``large enough'' number, $B$, and formulate the following constraints:

$$\forall a \in A:\quad \displaystyle\sum\limits_{l \in L} \omega_{a, l} * d_{a, l} \leq cp_a * B$$
$$\forall l \in L:\quad \displaystyle\sum\limits_{a \in A} \omega_{a, l} * d_{a, l} \leq cl_l * B$$

For our calculations, we set $B=10^{10}$ which is multiple orders of magnitude larger than the sum over all entries in the session matrix $S$. Using this ILP formulation, we can now calculate the maximum number of sessions that can be obtained for at most $P_{max}$ providers when connecting to at most $L_{max}$ peering LANs.

\subsection{ILP Formulation: Peering Scenario}\label{sec:theory:form:peering}

In our second scenario, we assume that the adversary chooses multiple settlement-free peers as injection ASes and then joins peering LANs to establish additional sessions with them. This case differs from the previous one, as routes are no longer propagated globally but rather only into the customer cone of the injection AS. We reuse the notation from §~\ref{sec:theory:form:transit}.

While we already defined the funneling degree, $FD_a$, of an AS $a \in A$ in §~\ref{sec:theory:assump}, we need to extend this concept to incorporate the number of sessions that can be established with the injection ASes. We can calculate the session-multiplied funneling degree (SMFD), $f^{P}_{a, l}$, for AS $a$ using only injection ASes in $I \subset A$ that are present at peering LAN $l$:

$$ f^{I}_{a, l} = \sum_{i \in I} \omega_{i, l} \cdot \mathbbm{1}_{PF_a}(i)$$
where $\mathbbm{1}_{Y}(x)$ represents the indicator function that returns 1 if $x \in Y$ and otherwise 0. 

\noindent\textbf{Parameters.} After calculating $f^{I}_{a, l}$ for each (peering LAN, ASN)-pair, we build the matrix $F$ as our first parameter: 

$$ F = 
 \begin{pmatrix}
  f^I_{a_1, l_1}  & f^I_{a_2, l_1}  & \cdots & f^I_{a_{|A|}, l_{1}}\\
  f^{I}_{a_1, l_2}  & f^{I}_{a_2, l_2}  & \cdots & f^{I}_{a_{|A|}, l_{2}} \\
  \vdots            & \vdots            & \ddots & \vdots  \\
  f^{I}_{a_1, l_{|L|}}  & f^{I}_{a_2, l_{|L|}}  & \cdots & f^{I}_{a_{|A|}, l_{|L|}} 
 \end{pmatrix}$$

We also provide the parameters $R \in \mathbb{N}$ and $N \in \mathbb{N}$ and a set of potential injection ASes, $I$. $R$ describes the required SMFD to count an AS as \textit{fully affected}, and $N$ describes the required number of fully affected ASes.

\noindent\textbf{Variables.} We add two binary decision variables, $d_l \in \{0,1\}, l \in L$ and $c_a \in \{0,1\}, a \in A$; $d_l$ determines whether the adversary should participate at peering LAN $l$ while $c_a$ tracks whether the current peering LAN selection introduce a session-multiplied funneling degree of at least $R$ for AS $a$.  

\noindent\textbf{ILP Problem Formulation.} Given $I$, $F$, $N$, and $R$, our goal is to minimize the resources---i.e., the number of peering LANs with which we have to establish a connection---needed to perform the \name attack, i.e., our objective function is:

\begin{equation*}
\begin{array}{lll}
\text{minimize} & \displaystyle\sum\limits_{l \in L} d_l &
\end{array}
\end{equation*}

Every valid solution should have a least $N$ fully affected ASes. Hence, we first add this constraint:
$$\displaystyle\sum\limits_{a \in A} c_{a} \geq N$$

Next, we want to assure that the combined SMFD (across all chosen LANs) of an AS is larger than $R$ for at least $N$ many ASes. Here, we utilize the fact that at least $N$ many $c_a$ variables are set to 1 (by the previous condition) while all other are set to 0. When we multiply $R$ by $c_a$ we effectively generate a switch that either does nothing or conditions the session-multiplied funneling degree of $a$ to be larger than $R$. As the described condition works only for a single AS, we have to add it once for each AS:
$$\forall a \in A:\quad \displaystyle\sum\limits_{l \in L} d_l * f^I_{a, l} \geq R c_{a}$$

Notably, this formulation does not incentivize the ILP solver to arrive at the solution with the largest number of set $c_a$ variables---each solution that sets at least $N$ of them is seen as equally good by the solver.

\subsection{Analysis \& Results}\label{sec:theory:analysis}

Now that we have formulated our two models, we can run an ILP solver with varying input parameters to explore \names cost-benefit trade-off landscape. 

\noindent\textbf{Implementation and Execution.} We implement the ILP program using Python3's PuLP library~\cite{pip2022pulp}. We configure PuLP to use the CBC C++ solver~\cite{git2022cbc} and time out (i.e., return the current best, potentially sub-optimal solution) after three hours. We refine sub-optimal solutions whenever possible, i.e., when an optimal run with stricter requirements produced a better objective value than a sub-optimal run, we copy the results from the optimal run over to the sub-optimal run.\footnote{e.g., when you need X peering LANs to affect 1000 ASes, you do not need more than X to affect 900 with otherwise identical configuration.}

\subsubsection{Transit Scenario}
\begin{figure}[h!tb]
    \centering
    \includegraphics[width=.66\linewidth]{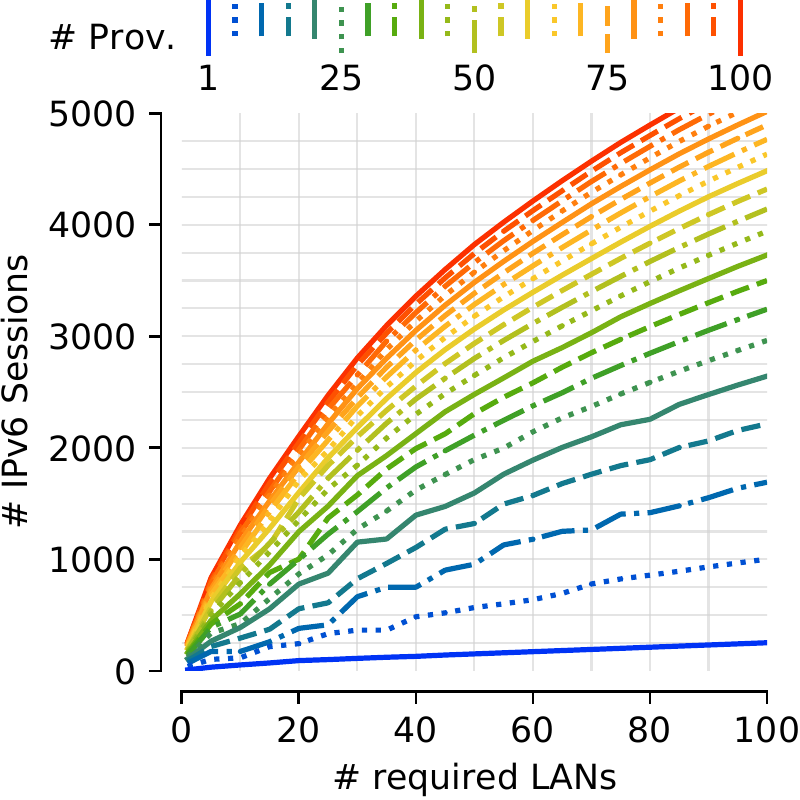}
    \caption{Transit Scenario: trade-off landscape.}
    \label{fig:theory:ana:transit}
\end{figure}
We solve the ILP problem defined in §~\ref{sec:theory:form:transit} for $L_{max}$ and $P_{max}$ values between 1 and 100 and obtain the maximum number of sessions that can be established using each pair. Figure~\ref{fig:theory:ana:transit} shows different lines for the number of transit providers ($P_{max}$), the number of peering LANs ($L_{max}$) on the x-axis, and the resulting number of obtainable sessions on the y-axis.

We first observe that we can establish more than a thousand transit sessions by choosing 20 providers and join 25 peering LANs. Given the many possibilities to remotely connect to a peering LAN as well as the cheap (in fact, often free) IPv6 transit options, deploying such an infrastructure is not a major hurdle. If each sessions allows us to send 1000 prefixes (which is not uncommon for transit sessions), this setup would already allow us to inject 1M routes into the global routing table. 

We further observe that we need to contract at least 35, 45, and 60 transit providers while joining at least 40, 60, and 80 peering LANs to establish 2000, 3000, and 4000 sessions via just a single port per peering LAN, respectively. While certainly harder to achieve, these scenarios are not out of reach for, e.g., state-backed adversaries.

\subsubsection{Peering Scenario}
\begin{figure*}[t!hb]
  \centering
  \includegraphics[width=\linewidth]{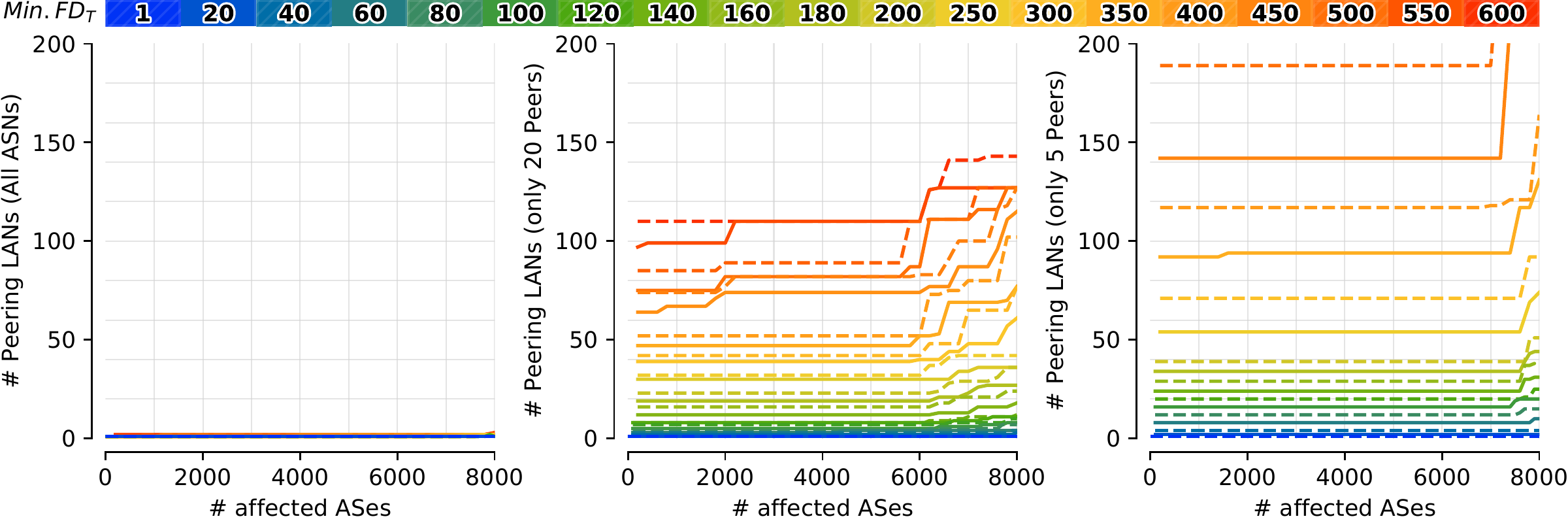}
  \caption{Peering Scenario: trade-off landscape for $I_{all}$ (left), $I_{20}$ (middle), and $I_5$ (right).}
  \label{fig:theory:ana:peering}
\end{figure*}

We solve the ILP problem defined in §~\ref{sec:theory:form:peering} for different required SMFDs ($R$), required fully affected ASes ($N$), and three different sets of injection ASes ($I$). We first choose $I_{all}$ to be the set of all IPv6-enabled ASes, which corresponds to setting up a bi-lateral peering link with each AS that participates at a peering LAN. While accomplishing this connectivity setup is unrealistic for new and small ASes, it provides us with a lower bound for the number of needed peering LANs. After that, we choose restricted sets of injection ASes, i.e., a scenario in which the adversary convinces a limited number of ASes to setup bi-lateral peering. In this scenario, choosing peers with large customer-cones and many sessions is the most ideal; hence, we rank ASes by the product of their customer-cone size and their total session count across all peering LANs and then choose the top 5 and top 20 ASes to represent the injection sets $I_5$ and $I_{20}$, respectively.

Figure~\ref{fig:theory:ana:peering} shows the resulting trade-off landscapes for $I_{all}$ (left), $I_{20}$ (middle), and $I_5$ (right). Each subplot shows the number of fully affected ASes ($N$) on the x-axis, different curves for the minimal required session-multiplied funneling degree ($R$), and the resulting minimal number of required peering LANs on the y-axis. The $I_{all}$ subplot shows that if an adversary could establish bi-lateral peering connections to all ASes at IXP LANs, connecting to a single (or few) peering LAN(s) is sufficient to generate $R=600$ for 8000 (and probably more) ASes. If the adversary can only establish peering with the injection ASes in $I_{20}$ or $I_5$, it is realistic to connect to enough peering LANs to introduce $R=200$ for 5000+ ASes, yet further increasing the required session-multiplied funneling degree might become a significant obstacle.

While a real adversary would realistically arrive at a setup somewhere between $I_{all}$ and $I_{20}$, properly representing the full spectrum of possibility, which is probably highly dependent on case-by-case, non-technical aspects (e.g., access to the right contacts, marketing, justification of need, ``prestige'' in the operator community, etc.), goes beyond the scope of this paper. Yet, our analysis shows that running \name \textit{solely} based on peering connections---which often have max-prefix limits of \textasciitilde100---seems unrealistic. This insight is further substantiated by our experiments in §~\ref{sec:exp:prop} which show that announcements via bi-lateral peering sessions do not necessarily propagate to all ASes within a peer's customer cone, which means that our calculated SMFDs are likely overestimates. 

\subsubsection{Discussion \& Feasibility}

While it is unlikely that an adversary acquires enough sessions via bi-lateral peering alone, we demonstrated it is possible to get thousands of sessions from various transit providers. Notably, our analysis took a very conservative approach for estimating the session count. In reality, an adversary could use 5, 10, or even more different VPS and remote peering providers simultaneously to establish multiple ports at each peering LAN, which would provide a linear multiplication factor to the number of sessions that can be established. Hence, a highly motivated adversary could potentially end up with 10k+ sessions, most of which capable to reach a significant portion of the IPv6 routing ecosystem. Even if each session would be tightly limited to 100 prefixes, such a setup could still produce an increase of 1M prefixes; hence, we conclude that performing \name is clearly feasible.

\section{Testing Router Behavior}
\label{sec:hwtest}

As the ``512k day'' in August 2014 (as well as its successors) received substantial media coverage~\cite{Emile2021512K,Emile2021768K}, router vendors are well aware of the possibility and potential impact of exceeding a router's available RIB or FIB memory. In this section, we examine how routers react to a large number of announced non-aggregatable IPv6 routes.

We perform our evaluation in our testbed with one popular enterprise router---the Juniper MX5 \cite{mx5}--- and one virtual version of a popular core router---the \cvr \cite{xrv9k}.
We use \exabgp \cite{exabgp}, a stateless BGP speaker, to quickly announce a large number of routes from a measurement machine to each of the two routers and assess the impact of those announcements over time.
We devise two different scenarios for our experiments: (1) the best-case scenario (from the victim's perspective), where each route contains the shortest possible AS path (i.e., a single AS, resulting in a path length of 1) and no BGP communities attached at all; (2) the worst-case scenario, where each route contains the longest possible AS path and maximum number of large BGP communities\footnote{The maximum possible AS path length and number of BGP communities that can be sent with ExaBGP is 251, even though the BGP \cite{rfc4271} and BGP large communities \cite{rfc8092} specifications allow even longer path attributes.}.
For both AS numbers as well as BGP communities we choose 32 bit values to maximize the impact on router memory.
For the hardware and the virtual router we use a minimal configuration whenever possible.
The Juniper MX5 does not have any prefix limit configured by default, while the \cvr has a default prefix limit of 524,288 for IPv6 \cite{cisco-prefix-limit}.
We increase XRv9k's prefix limits for our experiments.
Note that these prefix limits do not make \name infeasible (cf. \Cref{ssec:enablers}), in fact they can be easily circumvented by announcing prefixes over multiple sessions.
While we continuously announce new routes via ExaBGP, we monitor the resource usage of the system under test.

\subsection{Juniper MX5}

\begin{figure}[h!tb]
    \centering
    \includegraphics[width=.88\linewidth]{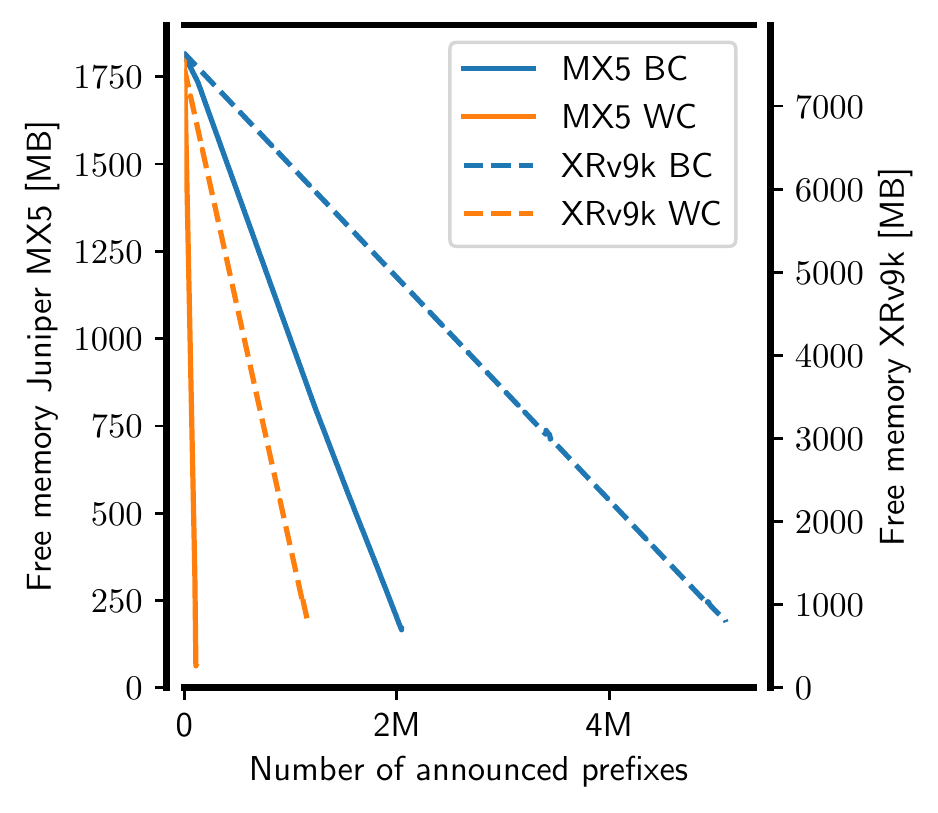}
    \caption{Juniper MX5 and Cisco XRv9k memory exhaustion for best-case (BC) and worst-case (WC) announcements.}
    \label{fig:router:combined}
\end{figure}

We begin our testbed experiments with the Juniper MX5 router.
In \Cref{fig:router:combined} we show the results of our memory exhaustion experiments.
In the best-case scenario, the router is able to accept around 2.04 million prefixes, before running out of memory.
In the worst-case scenario this number drops to a low 109k prefixes---which is substantially lower than the current number of all announced IPv6 prefixes (164k)~\cite{ipv6-tablesize}.

Once the router's memory is exhausted it will trigger an out-of-memory exception, which results in the BGP routing process being killed.
This results in a core dump of the routing process\footnote{Interestingly, the core file can be so large that it leads to the \texttt{/var} directory on the router becoming full, which can not be written to anymore, unless cleaned manually.}, a complete loss of all established BGP sessions, and a purge of all entries in the RIB and FIB.

\subsection{\cvr}
Next, we perform our experiments with the \cvr.
We show the results of our memory exhaustion experiments in \Cref{fig:router:combined}.
In the best-case scenario, the virtual router accepts slightly more than 5 million prefixes before running out of memory.
In the worst-case scenario, it only accepts around 1.16 million prefixes.

The virtual Cisco router deploys different levels of memory alerts \cite{low-memory}.
\one a \textit{minor alert} is triggered at 85\% memory occupancy which leads to rejection when trying to establish new eBGP sessions, whereas already established sessions are not affected.
\two a \textit{severe alert} is raised at 90\% memory usage and at that point the BGP daemon shuts down already established eBGP sessions until the memory threshold becomes minor.
The daemon shuts down BGP sessions with the lowest percentage of best paths selected (\# best paths from peer/\# total paths from peer).
\three a \textit{critical alert} will be triggered at 95\% memory usage, which leads to a shutdown of all established BGP sessions.
In our experiments we trigger all of these alerts sequentially, leading to a complete shutdown of all established BGP sessions.

\subsection{Theoretical Lower Bound Memory Usage}

We can also calculate the lower bound RIB memory usage of our worst-case announcements as follows:
\begin{equation*}
\begin{array}{ll}
MEM = & (PREFIX\_SIZE + (255 \times ASN\_SIZE) + \\
& (255 \times COMM\_SIZE) ) \times NUM\_PFX 
\end{array}
\end{equation*}

Assuming a prefix size of 16 bytes for the IPv6 prefix and 1 byte for the IPv6 prefix length, an ASN size of 4 bytes, and a BGP large community size of 12 bytes, we get a lower bound of $ MEM = 4097 \times NUM\_PFX$, i.e. every worst-case prefix needs at least 4kB of RIB memory.
Given that today's core routers (\eg Cisco ASR 9000, Juniper MX960, or Arista 7280CR2K) have RIB memory sizes of 32 or 64 GB, a large number of worst-case prefixes can still bring a router with lots of memory to its knees: 8M prefixes---which can be obtained from de-aggregating a /26---suffice to fill up 32 GB.
Finally, as the IPv4 Internet is approaching the 1M route threshold \cite{cidrv4} and the increasing deployment of technique such as RPKI \cite{rpkiblog}, fewer routes from an attacker will be needed to further fill up a router's RIB memory.

\takeaway{Enterprise routers can already be overwhelmed with as little as $\approx$ 100k announcements, whereas core routers can at least handle around 1M. In the worst case, a route needs at least 4KB router memory to be stored.}

\section{Real-world Experiments}
\label{sec:experiments}
Due to ethical concerns as well as economical and social consequences, we can not simply perform a large scale attack on the Internet to provide a proof-of-concept. Instead, we opt for multiple small-scale experiments that provide interlocking insights into the viability of different parts of the attack. 

\subsection{Obtaining Resources and Connectivity} \label{sec:exp:resources}
We state in Section \ref{ssec:enablers} that it is fairly easy to \one receive the resources needed to execute the proposed attack, \two join multiple IXP peering LANs, and \three establish additional sessions to large transit providers. Below we report on our experience in building and operating a proof-of-concept network capable of performing a small-scale \name attack at negligible cost. 

\noindent\textbf{Internet Resources.} We obtained an AS number (\dbsub{AS39282}) and a few IPv6 address blocks (\dbsub{2a10:cc47:100::/40, 2a0e:b107:e80::/44, and 2a10:2f00:15d::/48}) through a sponsoring LIR (Securebit), at a total cost of 270 EUR (valid for 1 year).
It took only a few days from requesting these resources until obtaining them for use on the Internet.

\takeaway{It is possible to obtain ASNs and IP prefixes in a matter of days and at cost bearable for individuals.}

\noindent\textbf{Peering LANs.} We built our proof-of-concept network using 2 VMs with IXP access: one in Frankfurt (provided by vServer.site) and another in Dusseldorf (provided by Securebit).
This allowed us to directly access all route servers and peering LANs of 4 medium-to-large size IXPs: DEC-IX, NL-IX, KleyReX, and LocIX.
In total, we paid an initial setup fee of 160 EUR and a monthly operating fee of 60 EUR. 
It took a day till we connected to the first IXP and a few weeks until we connected to the last IXP.

\takeaway{IXP connectivity providers let new ASes quickly join many peering LANs at a small cost.}

\noindent\textbf{Transit Sessions.} We decided to use Hurricane Electric (HE, AS6939) as our main transit provider, as it is one of the most important IPv6 networks~\cite{jia2019tracking}.
Surprisingly, HE reached out to us about setting up bilateral peering sessions at our IXPs---with a free IPv6 transit option---before we even knew the IXP on-boarding process finished.
Additionally, we obtained a VM in Amsterdam from Vultr (AS20473), which provides BGP transit to its customers at no additional cost.
We paid no setup fee and a monthly operating fee of 5 USD.
The VM was available in a few minutes.

\takeaway{It is possible to instantly get cheap IP transit.}

\noindent\textbf{Prefix Limits.} After finding out our sessions have low prefix limits, we asked if our providers could raise them.
As a result, in less than 24h, most operators increased the limits by an order of magnitude without asking for explanation.
Other operators stated they could arbitrarily raise the limits given a reasonable justification.

\takeaway{Increasing prefix limits is a matter of asking, and often requires no justification.}

\begin{figure*}[t!hb]
        \minipage[t]{0.32\textwidth}
          \includegraphics[width=\linewidth]{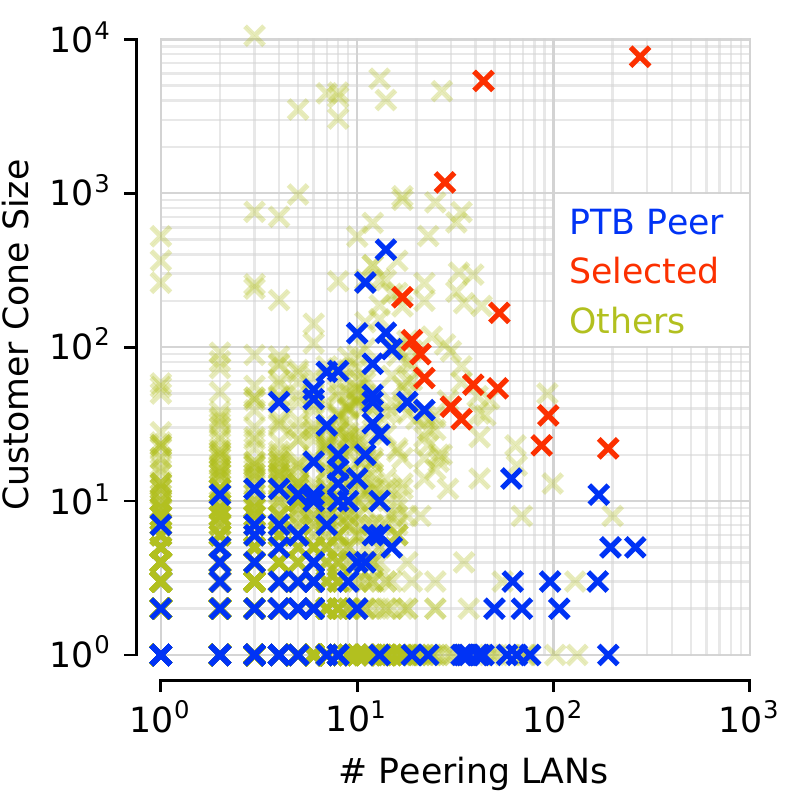}
          \caption{PEERING testbed peers: customer cone vs. peering LANs.}
          \label{fig:tested_asns}
        \endminipage
    \hfill
        \minipage[t]{0.32\textwidth}
          \includegraphics[width=\linewidth]{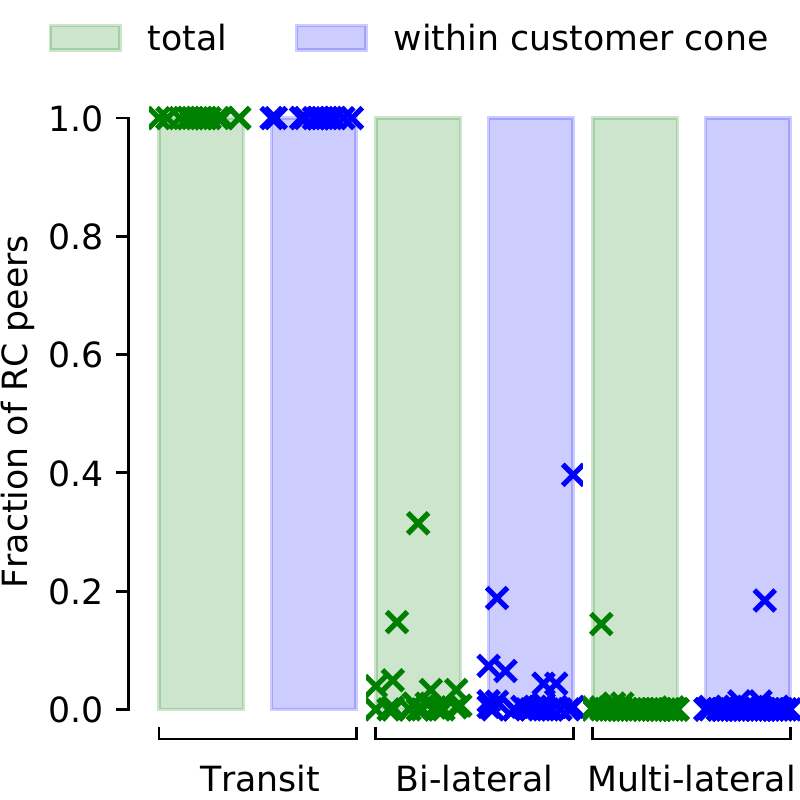}
          \caption{Redistribution behavior of different session types.}
          \label{fig:redistr_exp}
        \endminipage
    \hfill
        \minipage[t]{0.32\textwidth}
          \includegraphics[width=\linewidth]{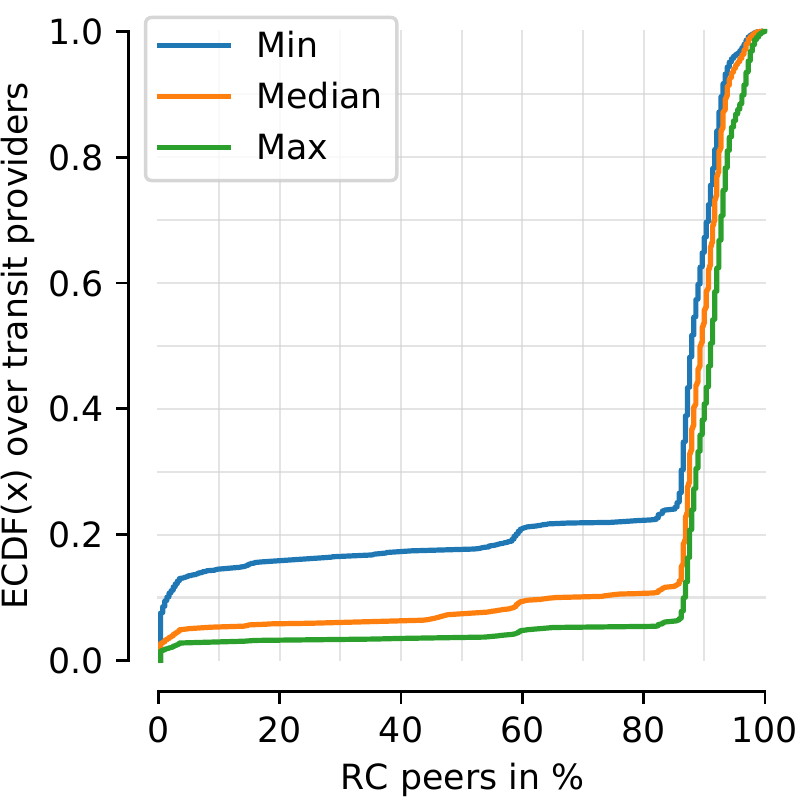}
          \caption{Redistribution behavior for transit providers of single-homed ASes.}
          \label{fig:redistr_ana}
        \endminipage
\end{figure*}

\subsection{Propagating Announcements}\label{sec:exp:redistribution}
Below we take a closer look at the routing ecosystem itself. In particular, we analyze the correctness of the claims we made earlier in Sections \ref{sec:overview} and \ref{sec:theory}. We use the infrastructure described in the previous subsection and the PEERING testbed to run real-world experiments for a limited number of ASes and contrast our findings with insights obtained from the routing information captured by the route collector projects. 

\subsubsection{Setup Specifications} 

We make use of the proof-of-concept network that we built in the previous subsection to produce IPv6 route announcements. Besides the thousands of (implicitly gained) multilateral peering sessions via route servers, our network only has few direct sessions (most of which connect to HE). To improve our coverage of large IPv6 transit providers and, thereby, improving the generalizability of our results, we also utilize the PEERING testbed~\cite{schlinker2014peering,schlinker2019peering}. The PEERING testbed is a research network that allocates resources (i.e., ASNs and prefixes) to submitted and accepted project proposals. It has 207 direct IPv6 sessions to 150 different networks distributed across 9 physical locations as well as dedicated IPv6 sessions to 12 route servers at 5 IXPs. All announcements from the PEERING testbed were originated from AS 47065 and sourced from the \dbsub{2804:269c:10::/44} IPv6 address block. In addition to the standard project capabilities we received the additional capability to announce BGP communities that control the redistribution behavior of the connected route servers.

\noindent\textbf{Announcement Schedules.} We announced a dedicated /48 IPv6 prefix via each session. As we control fewer unique /48 prefixes than we have sessions, we first organize the sessions into groups and then reuse the same prefixes across groups (but not within each group). To substantially reduce the likelihood that two successive groups are influenced by one another (e.g., as the first one triggers Route Flat Damping), we adopt a two hour announcement schedule---we announce all prefixes within a group, then wait 30 minutes for route convergence, then withdraw all prefixes, and then wait another 90 minutes before repeating the cycle with the next group. While, e.g., MRAI timers~\cite{gill2012effect} or similar update minimization techniques may introduce few minutes of delay to the propagation of our our announcements, we have to wait additional 60 minutes in the last step to ensure that accidentally triggered Route Flap Damping penalties expire~\cite{gray2020bgp} and can hence no longer influence the next group of announcements. 

\noindent\textbf{Routing Information.} We utilize the route collector projects RIPE RIS and Routeviews as our vantage points. In total, they operate 47 IPv6-enabled route collectors that connect to 305 full-feed ASes via 555 IPv6 sessions. For our analysis, we utilize all available RIB snapshots at 2022-09-26, 00:00 UTC+0 using the BGPStream utility.

\renewcommand{\arraystretch}{0.9}
\begin{table}[h!tb]
    \centering
    \begin{tabular}{@{}llll@{}}
    \toprule
                                & Routes & Paths & Prefixes\\
        \midrule
Total & 58.2M & 13.9M & 223K\\
AS set & 12K (0\%) & 10K (0\%) & 57 (0\%)\\
ATOM. & 4.2M (7\%) & 1.0M (7\%) & 161K (72\%)\\
AGGR. & 5.1M (8\%) & 1.3M (9\%) & 16K (6\%)\\
Any Hint & 6.4M (10\%) & 1.6K (11\%) & 162K (72\%)\\
    \bottomrule
    \end{tabular}
    \caption{Results of aggregation analysis.}
    \label{tab:routeagg:res}
    \vspace{-2.5em}
\end{table}

\subsubsection{Route Aggregation} \label{ssec:route-aggregation}

In this first experiment, we announce pairs of aggregatable routes via all our transit providers, i.e., HE at our infrastructure and 7 different transit providers at the PEERING testbed. We repeat this experiment twice. The first time, we announce two consecutive prefixes (i.e., A:B:C::/48 and A:B:C+1::/48) via each session. As both routes are entirely identical, a transit network may decide to aggregate these two routes and only redistribute the resulting /47 route that covers both announcements. The second time, we announce a /47 covering prefix and the /48 sub-prefix with the same network address (i.e., we announce A:B:C::/47 and A:B:C::/48 but not A:B:C+1::/48). In this scenario, a transit AS may decide to not redistribute the more-specific /48 route given that the AS path is identical. While we see all announcements propagate globally (i.e., each prefix is seen by at least 95\% of all route collector peers), we see no signs of aggregation.

\noindent\textbf{Analysis.} When an AS aggregates a route, it may leave up to three clues in the BGP messages that it redistributes. First of all, AS paths may consist of AS sequences and AS sets~\cite{rfc4271}. A set is generated whenever two routes with different AS paths are aggregated; they represent a summary of the non-matching parts of the two initial AS paths. If an AS aggregates a route and generates no AS set during this process, it should add the \texttt{ATOMIC\_AGGREGATE} attribute to the message. Finally, an AS may set the \texttt{AGGREGATOR} field to indicate that it produced this route aggregate. We searched all IPv6 routes seen by the route collectors for these three hints and display our findings in Table \ref{tab:routeagg:res}. While we observe that 72~\% of prefixes have at least one path with an aggregation hint, we only observe 11~\% of paths and 10~\% of routes with aggregation hints; hence, we believe that only few ASes actively perform route aggregation. While we did not find any signs of route aggregation during our own experiments, an adversary could also make routes less aggregatable by announcing neither neighboring nor covering prefixes to the same neighbor, and alternating the origin AS. 

\takeaway{While aggregation is a theoretical challenge, it is rare in practice and can be circumvented.}

\subsubsection{Route Redistribution}\label{sec:exp:prop}

Next, we want to analyze whether our assumptions for the route propagation behavior of transit, bi-lateral, and multi-lateral sessions are accurate. While the number of transit providers for both testbeds is limited, applying our schedule to all bi-lateral and multi-lateral peers connected to the PEERING testbed would require extensive amounts of time; hence, we select a smaller set of important ASes. 

\noindent\textbf{Tested Networks.} The importance of a network for our attack can be characterized by two dimensions: the number of sessions we can establish with it, and the number of networks it redistributes our announcements to. Figure \ref{fig:tested_asns} shows the customer cone size (y-axis) against the number of peering LANs to which a network is connected (x-axis) as a scatter plot for all networks with PeeringDB entries. We mark networks that connect to the PEERING testbed in blue (``PTB Peer'') or red (``Selected'') and all other networks in green (``Others''). As both dimensions are equally important to \name, we select the 15 PEERING peers with the highest harmonic mean\footnote{Compared with arithmetic mean, the harmonic mean leans towards lower numbers, which penalizes networks that appear large in only one dimension.} of customer cone size and the number of potential sessions.

\noindent\textbf{Experiment.} Figure~\ref{fig:redistr_exp} shows the fraction of route collector peers (y-axis) reached by /48 announcements via each of the three different session types (on the x-axis). We calculate this fraction twice: once relative to all IPv6 route collector peers (green, ``total'') and once relative to the peers within the customer cone of the neighbor to which we announced the prefix (blue, ``within customer cone''). We can first verify that announcements towards transit providers always propagated globally and that announcements via multi-lateral peers barely propagates at all. Yet, contrary to our assumption, not a single bi-lateral peering sessions redistributed our announcements into even half of its customer cone. Hence, we likely over-estimated the achievable funneling degrees in §~\ref{sec:theory:form:peering}, which we already noted in that section. 

\noindent\textbf{Analysis.} To further test the validity of our transit propagation assumption, we analyze the public BGP data. After removing path-prepending~\cite{marcos2020path}, we select all prefixes for which all paths have the same first-hop AS, i.e., that were announced via a single transit provider. Figure~\ref{fig:redistr_ana} shows the minimal, median, and maximal propagating route for each of these transit providers as an ECDF. We observe that for 80~\% of transit providers every route propagates globally (i.e., to more than 80~\% of route collector peers), while for 89~\% and 94~\% at least the median and best route propagated globally, respectively.  

\takeaway{While bi- and multi-lateral peers do not necessary redistribute into their entire customer cones, announcing to a transit provider leads to global redistribution.}

\section{Discussion}
\label{sec:discussion}

\textbf{Targetability \& Collateral damage.} While we introduced \name as a global attack, BGP has many mechanisms that allow an adversary to steer the redistribution of a route. Many transit providers allow their customers to directly decide to which neighbors their routes get redistributed by attaching specifically encoded BGP community attributes~\cite{RFC1997,streibelt2018bgp,birge2019sico}. In addition to this collaboration-based technique, the adversary may also ``poison'' the AS path to avoid that certain ASes accept the route. The poisoning method leverages cyclic route filters implemented by most routers: if the adversary $A$ forges a route with the path $A X A$ and this route propagates to AS $X$, $X$ will likely drop it~\cite{katz2011machiavellian}. As the Internet's routing hierarchy has flattened drastically over the last decade, it is likely that a combination of these two mechanisms could be sufficient to steer routes towards most regional networks. Yet, even if the adversary succeeds in steering the majority of the attack towards a single AS, the increase in redistributed routes at the intermediate ASes should still be very noticeable providing an opportunity to detect the attack and limit the redistribution. 

\noindent\textbf{Detection \& Mitigation.} \name is easily detectable as it introduces multiple times more routes than the current IPv6 routing table contains in total.
Hence, even operators that do not monitor their own network could detect it by checking Twitter notifications from the IPv6 routing table size bot or the BGPStream bot~\cite{CISCO2022BGPstream,TWITTER2022BGP6}. Once some operator detected the attack and shared the origin ASes involved in it via some high-visibility operator mailing list such as NANOG, the attack can be mitigated by adding simple ingress filters for the covering prefix and origin ASes. Once these filters are added, the routers no longer import any routes related to the attack which should prevent the router from running out of memory and also drastically lower the CPU load. Due to the simplicity of the mitigation process, \names attack duration is effectively limited to how quickly network operators (especially those of intermediate ASes) can co-ordinate the mitigation efforts---a time that we hope to substantially reduce by raising awareness via this paper and our carefully designed disclosure process. 

\noindent\textbf{Traceability \& Repercussions.} \names resources are easily traceable to the RIRs that allocated them and, from there, might be directly accountable to a specific person or organization. While this seems like a large issue, there are no real sanctions or direct repercussions for ``routing vandalism.'' Bitcanal illustrates this issue nicely: besides loosing some ``reputation'' via call-outs from researchers and operators~\cite{SecBoul2022Bitcanal,testart2019profiling}, Bitcanal continued to frequently hijack the resources of other ASes over multiple years, until Spamhaus added all related ASNs to their ``Don't route and peer'' list~\cite{SPAMHAUS2022DROP}.

\subsection{Potential Defense Mechanisms}
While \name can be mitigated quickly, we ideally would like to entirely prevent it from being feasible. Yet, based on its distributed nature, there is no simple solution that fully prevents the attack; however, there are multiple technical and non-technical mechanisms that may limit \names impact or increase its requirements. 

\noindent\textbf{Dynamic Yet Tight \texttt{Max-Prefix} Limits.} Transit providers should introduce dynamically growing yet tight per-session limits on their eBGP sessions. We recommend to allow customers and peers to announce at most 1.5x the number of prefixes they announced the previous day. Similarly, the IPv6 routing table currently grows <50k new prefixes per year~\cite{ipv6-tablesize}; hence, we further recommend to allow a maximum daily increase of at most few thousand prefixes on transit sessions. Automatic imports of max-prefix limits from, e.g., PeeringDB should be sanity checked and not be allowed to surpass a certain pre-defined limit---otherwise adversaries could enter arbitrary high numbers and abuse the prevention automation. 

\noindent\textbf{Per-Origin and Per-Block Prefix Limits.} We highly recommend transit providers to stop the redistribution once too many routes within the same covering prefix or by the same origin AS are announced. While covering prefixes would optimally be determined by analyzing the daily IRR delegation files, counting on a /29 or /32 basis might be easier to implement. Currently, the AS with most announcements is AS9808 with 3870 IPv6 routes and the covering prefix with the most more-specific announcements is 2409:8000::/20 with 9807 more-specifics. As implementing these limits on each router is costly (and may still be insufficient if different routers receive unique sets of routes), we highly recommend (if available) to introduce these limits on a route reflector.

\noindent\textbf{Tight Resource Monitoring \& Filtering.} We recommend transit providers to monitor the number of sessions that other ASes establish with them---especially if their peering policy is fully open or they employ a fully automated session establishment service. If they automatically generate filter lists from third party data sources (e.g., RPKI~\cite{Manrs2022rpki}, IRR~\cite{Merit2022IRR}, or Team CYMRU~\cite{Cymru2022Bogon}), we recommend them to carefully monitor the resulting filter size; checking the number of acceptable prefixes may reveal the preparation for a \name attack early. Further, we recommend transit providers to only redistribute what is correctly registered and avoid loose filtering, i.e., do not assume that more-specific versions of route objects or ROA records are valid by default. While this will not directly prevent the attack, it will increase the effort on the adversary's side to register the resources correctly.

\noindent\textbf{Delayed Propagation of Unfamiliar Routes.} The concept behind Pretty Good BGP \cite{karlin2006pretty} is to avoid propagation of anomalous routes, not seen in a window of historical data. Thus, the use of previously unseen routes is delayed, with the hope of identifying and neutralizing any attacks in the meantime, if the route was really malicious. In the context of \name, if the attacker used a hijacked prefix, the idea presented in that paper could stop the attack from propagating further, yet note that prefix history tracking needs memory anyway. On the other hand, as \name does not need IP prefix hijacking, the attacker can use a large pool of addresses that has never been announced on the Internet before. Thus, we suggest modifying the Pretty Good BGP idea so that it \emph{also} delays accepting announcements of new prefixes that are not contained in already propagated, larger address blocks. %
\section{Ethical Considerations}
\label{sec:ethics}
The attack framework we present in Section~\ref{sec:overview} has the potential to cause serious harm.
Hence, our research naturally raises several ethical concerns, which we discuss in this section.

\textbf{Real-world Experiments.} While we performed a thorough theoretical evaluation of our attack's potential impact (Section~\ref{sec:theory}) and assess the behavior of various hardware implementations exclusively in a non-Internet-connected lab environment (Section~\ref{sec:hwtest}), the Internet is a dynamic system, and---given the issue of deaggregation is well known---might not be susceptible to the attack after all.
Hence, we also conducted real-world experiments, see Section~\ref{sec:experiments}.

When designing our experiments, we closely followed the Menlo report \cite{kenneally2012menlo} and work by Partridge and Allman \cite{partridge2016ethical} to mitigate potential harm to the Internet.
This includes a thorough harm benefit analysis after assessing the theoretically possible impact.
Here, we weighted that the underlying techniques of our attack are generally known in the community, and it is likely that others with potentially malicious intentions may independently develop our scaling methods for prefix deaggregation. 
At the same time, the networking community considers existing techniques---like per session prefix limits---sufficient to mitigate the threat, and is unlikely to consider our attack as serious and implement prevention mechanisms, unless the feasibility can be practically demonstrated.

Hence, we decided to conduct a small-scale experiment using 500 prefixes via Vultr. Given the size of the IPv6 routing table (\textasciitilde160k prefixes), we believe that that these 500 prefixes (\textasciitilde0.3\%) were well inside the daily BGP IPv6 table size churn.
Furthermore, we limit the duration of our announcements, made them unlikely to trigger route flapping, and ensured that our announcements are properly withdrawn after we have completed the experiments.
In our PEERING testbed experiments we never announced more than 20 IPv6 prefixes simultaneously. 
Similarly as with Vultr, we ensure proper withdrawal of our announcements in the PEERING testbed.

\noindent\textbf{Independent Reproduction by Unknown 3rd Party.} Despite our best efforts to design an experiment that does not cause harm, it was still visible in the global routing table.
Six days after we conducted our experiments---which did not cause noticeable load at an independent leaf AS we operate as well---we observed an unknown entity that replicated our experimental setup executing \name with over 8,000 prefixes from one /32 via Vultr.
This caused noticeable load on the independent leaf AS we operate and was widely noticed in the operator community.

We hence decided to accelerate the initial disclosure process we had planned to take place (see below).
Furthermore, it demonstrates that, by now, threat actors are actively monitoring the global routing table.
Researchers conducting experiments for potential vulnerabilities in the routing ecosystem \emph{must} consider that even small-scale experiments may reveal attack opportunities to third parties.
This leads to substantial problems when the ``attack'' opportunity is (technically) well-known in the community, yet is currently not considered ``exploitable enough''
\cite{dietrich2018investigating}.
\noindent\textbf{Disclosure Schedule.} After an independent third part potentially replicated our experiments on a significantly larger scale, we immediately launched a two-stage notification process. While technically, a lengthy coordinated vulnerability disclosure process~\cite{householder2017cert} would have been preferred, also to have more time to carefully discuss with operators \emph{why} this well-known vector is a higher threat \emph{now}, we opted for this path due to the actions of the unknown third party around the 5\(^{th}\) of October, 2022~\cite{Vultr2022event}.

\begin{itemize}[leftmargin=*]
\item \textbf{Private Disclosure Phase (2022/10/11--19).} We first disclosed the details of our attack via a whisper-network of well connected Tier-1 network operators and IXPs. In this process, we distributed the document enclosed in \Cref{fig:disclosure:first}. This process included 8 major IXPs, 20 Tier-1 ASes, and 7 major content providers. %
We followed-up the initial notification with a clarifying statement,
highlighting that an independent third party potentially already executed the attack on a larger scale.
We received the feedback that this clarification made the severity of the problem apparent.

\item \textbf{Public Disclosure Phase (2022/10/20 and onward).} After sufficient reaction time and no signals to further delay the disclosure, we publicly disclosed our findings via 13 different operator mailing lists (including, e.g., NANOG, DENOG, and the RIPE Routing Working Group) as well as via different social media platforms. 
\end{itemize}

During our disclosure phases, we continuously discussed our findings with network operators, integrated their experiences, and assisted them in deploying prevention mechanisms whenever possible. From private e-mail exchanges, we know that at least two Tier-1 ASes, three cloud providers, and various smaller networks actively configured prevention mechanisms against our new form of prefix de-aggregation attacks. 
\section{Summary}

In this paper we presented \name, an attack that overwhelms BGP routers by globally distributing millions of IPv6 routes via thousands of distributed sessions. We demonstrated that \name can bypasses traditional prevention mechanisms via its distributed nature and showed that its required infrastructure and resources can be obtained swiftly and at a cost bearable even for single individuals. We tested our assumptions in lab experiments, real-world measurements, and by analyzing passively obtained routing information. Finally, we launched a two-stage disclosure campaign to notify network operators and expedite the deployment of prevention mechanisms.  
\bibliographystyle{IEEEtranS}
\bibliography{bdti}
\balance

\appendix
\section{Private Disclosure Notification}\label{sec:app:discl:ini}

\Cref{fig:disclosure:first} shows the initial email that we sent out in the private disclosure notification. In \Cref{fig:disclosure:second} we show the follow-up email highlighting why the attack can cause serious harm and has already been run on a larger-scale by an unknown third party.

\lstset{
  basicstyle=\linespread{0.8}\scriptsize\ttfamily,
  columns=fullflexible,
  keepspaces=true,
  breaklines=true,
  breakindent=0pt,
  frame=single,
}

\begin{figure}[h!]
\begin{lstlisting}
Dear colleagues,

we received some feedback that the message we provided you with is simply stating the obvious, and noticed an important piece of information missing:

Note, that we conducted experiments with a limited (<=500 prefixes) test-setup around the 29th of September. On the 5th of October an entity unknown to us replicated our experiments via AS20473 with around 8k prefixes, already causing noticeable load but yet staying below the potential of this technique. We hence assume that our technique is by now known--not only commonly known in the community but potential attackers being consciously aware--to third parties, which is why we are sending out these notifications for something technically well known. We plan to notify the wider networking community in one week from now.

With best regards,
<blanked>
\end{lstlisting}
\caption{Follow-up email text of private disclosure notification.}
\label{fig:disclosure:second}
\end{figure}

\begin{figure*}[h!tb]
\begin{lstlisting}
Dear <Person>,

I'm a researcher at the Max Planck Institute for Informatics in Germany and received your contact from <Person>, who believes that you might be the right contact at <Company> for the following issue:

We started the private disclosure process for an IPv6-based routing attack discovered in a research collaboration between the Max Planck Institute for Informatics in Germany and the Institute of Theoretical and Applied Informatics, Polish Academy of Sciences. We'd highly appreciate your valuable insights and hope you join our efforts in globally deploying effective prevention mechanisms. To keep the Internet and its users safe, it is important to keep the attack details confidential until prevention mechanisms are in place; we count on you not to publicly share this information prior to the public disclosure, which we currently plan for Wednesday, 19th October 2022.

# What is the problem?

Routers either crash, drop sessions, or behave in other unintended ways when their FIB or RIB runs out of memory. While newer routers can store up to 4M prefixes, many ASes still run (at least some) older hardware that may only be able to store 1M routes or even less. TL;DR: We found an attack that allows an adversary to introduce very quickly more than 1M new and unique IPv6 prefixes into the global routing table and is only preventable with the help of major transit networks and IXPs. If, afterwards, these prefixes also get withdrawn simultaneously, the resulting path-hunting behavior additionally results in a massive flooding attack.

# How does the adversary even obtain 1M unique prefixes?

After obtaining a /29 address block from any of the RIRs (e.g., RIPE this does not even require need-based justification) the adversary announces every possible /48, /47,... /29 route leading to the announcement of 1.048.575 unique routes---if C is the difference between the minimal propagating CIDR size, /48, and the CIDR size of the address block from which an attacker sources routes, the adversary can announce up to 2^(C+1) - 1 unique routes, e.g., a /46 block can source seven routes in total: one /46 route, two /47 routes, and four /48 routes.

# Don't we have per-session prefix limits that prevent such attacks?

If the average per-session limit is X, an adversary 'simply' has to distribute its routes via 1M/X many sessions, i.e., per-session limits do not eliminate the issue, they only transform it into a session-hunting challenge. During our real-world experiments and discussions, we noticed that while many ASes set tight (often 100-500 prefixes) per-session limits on their peering sessions, it's less common that ASes on either side of a transit session enforce prefix limits.

# Why does ROV not protect us from this attack?

It is possible to set a single ROA entry that specifies that the /29 prefix can be announced with CIDR sizes up to /48. If the adversary generates such a ROA and waits some days for it to propagate to all validating ASes, each of the more than 1M prefixes would be a valid announcement.

# How can an adversary even get hundreds or thousands of sessions?

The idea is that remote peering providers and VPS providers (e.g., Vultr) enable the adversary to quickly and cheaply 'click together' (virtual) ports at many (think 20+) different peering LANs. The adversary obtains transit by picking providers that also establish transit sessions over peering LANs (Hurricane Electric being the prime example), many bi-lateral peering sessions via openly/aggressively peering networks (that can be identified via, e.g., PeeringDB), and additional (less effective) sessions via multi-lateral peering with Route Servers. Surprisingly, while it would be hard to assemble enough sessions with just one port at each peering LAN (yet eventually doable), this limitation does not exist in reality; while certain providers directly allow clicking multiple ports for a single peering LAN, there are also multiple providers---this allows the adversary to obtain a 5X to 10X factor for its session counts by establishing multiple sessions to each neighbor (in fact each port of each neighbor).

# Do these routes even propagate far enough?

TL;DR: yes. As a rule of thumb: The routes announced via transit sessions usually propagate globally, routes announced to bi-lateral peers usually propagate into the peer's customer cone, and routes announced via multi-lateral peering usually propagate only to the peer's regional customers. As part of our research, we analyzed the propagation behavior and found that an adversary that combines announcements via all three peering types can inject lethal amounts of IPv6 routes into routers of 8k+ ASes, i.e., yes, enough of these routes propagate far enough.

# Don't ASes along the path aggregate the individual routes?

While some ASes do aggregate routes, it is possible to launch the attack in such a way that routes can not be aggregated: the adversary would have to choose the prefixes in each session in such a way that neither two consecutive prefixes nor a prefix and its covering prefix are announced via the same session and/or neighbor. To be extra safe, the adversary could switch between multiple origin ASNs for the announcements or use path-poisoning to alter a route's AS path.

# What can IXPs do to help prevent the attack?

Ensure that your route servers have tight prefix limits and that they only accept a small number of sessions from each participant.

If applicable, monitor your members' session acquisition behavior (e.g., by looking for BGP-session related packets in the peering LAN's traffic data) to identify potential adversaries early.

# What can transit providers do to help prevent the attack?

Introduce dynamically growing yet tight per-session limits on all of your sessions. Allow, e.g., customers and peers to announce at most 1.3x the number of prefixes they announced yesterday. Similarly, the IPv6 routing table currently grows at a rate of <50k new prefixes per year; hence, one could limit the maximum daily growth to, e.g., at most 10k prefixes.

Closely monitor the number of sessions that other ASes establish with you---especially if your peering policy is fully open or you employ a fully automated session establishing service.

Given that the attack model is highly distributed, the best position to install protection mechanisms is your route reflectors, as they often have a complete view of the globally redistributed routes. If possible, implement the following two limiters:

(i) ensure that you only accept and redistribute a certain number of routes per origin AS

(ii) ensure that you only accept and redistribute a certain number of more-specific routes for each assigned address block.

(iii) accept only what is correctly registered. Do not allow an automatic "or longer" for any registered prefix. This will not prevent the attack but add more effort on the attackers' side to register the resources correctly.

(iv) monitor your generated filter size. A simple check on the number of acceptable prefixes can reveal the preparation of such an attack.

If you have any further questions, please don't hesitate to contact me!

Best regards,

<blanked>
\end{lstlisting}
\caption{Private disclosure notification email text.}
\label{fig:disclosure:first}
\end{figure*}

 \end{document}